\documentclass[12pt,a4paper]{article}
\usepackage[latin9]{inputenc}
\usepackage{textcomp}
\usepackage{amsmath}
\usepackage{amssymb}

\makeatletter

\newcommand{\lyxaddress}[1]{
	\par {\raggedright #1
	\vspace{1.4em}
	\noindent\par}
}

\@ifundefined{date}{}{\date{}}

\usepackage{amsfonts}
\usepackage{amsthm}
\usepackage{mathrsfs}

\renewcommand\@biblabel[1]{$^{#1}$} 

\renewcommand\maketitle
  {\noindent
   {\Large\bfseries\@title}%
   \vskip\baselineskip\par
   {\hskip1.5em\@author}%
   \hfill
   {\@date}%
   \bigskip\par\noindent
  }

\newtheorem{theorem}{Theorem}
\newtheorem{proposition}[theorem]{Proposition}
\newtheorem{lemma}[theorem]{Lemma}
\newtheorem{corollary}[theorem]{Corollary}
\newtheorem*{lemma*}{Lemma}

\theoremstyle{remark}
\newtheorem{remark}[theorem]{Remark}
\newtheorem*{remark*}{Remark}
\newtheorem*{remarks*}{Remarks}
\newtheorem*{example*}{Example}
\newtheorem*{question*}{QUESTION}
\newtheorem*{conjecture*}{CONJECTURE}

\theoremstyle{definition}

\newtheorem*{definition*}{Definition}

\newtheorem*{notation*}{Notation}

\renewcommand{\Re}{\mathop\mathrm{Re}\nolimits}

\makeatother

\begin{document}
\title{Coulomb Green's function and an addition formula\\
\vskip-0.5em\noindent for the Whittaker functions}
\author{Pavel \v{S}\v{t}ov\'\i\v{c}ek}
\maketitle

\lyxaddress{\hskip2.6em\emph{ Department of Mathematics, Faculty of Nuclear
Science, Czech Technical}\\
\hskip2.9em \emph{University in Prague, Trojanova 13, 120~00 Praha,
Czech Republic}}
\begin{abstract}
\noindent A series of the form $\sum_{\ell=0}^{\infty}c(\kappa,\ell)\,M_{\kappa,\ell+1/2}(r_{0})W_{\kappa,\ell+1/2}(r)P_{\ell}(\cos(\gamma))$
is evaluated explicitly where $c(\kappa,\ell)$ are suitable complex
coefficients, $M_{\kappa,\mu}$ and $W_{\kappa,\mu}$ are the Whittaker
functions, $P_{\ell}$ are the Legendre polynomials, $r_{0}<r$ are
radial variables, $\gamma$ is an angle and $\kappa$ is a complex
parameter. The sum depends, as far as the radial variables and the
angle are concerned, on their combinations $r+r_{0}$ and $(r^{2}+r_{0}^{\,2}-2rr_{0}\cos(\gamma))^{1/2}$.
This addition formula generalizes in some respect Gegenbauer's Addition
Theorem and follows rather straightforwardly from some already known
results, particularly from Hostler's formula for Coulomb Green's function.
In addition, several complementary summation formulas are derived.
They suggest that a further extension of this addition formula may
be possible.\\
\end{abstract}

\section{{\large{}INTRODUCTION}\label{sec:Introduction}}

Green's function of a Hamiltonian is an important object in quantum
physics as it contains, in principal, all information about the respective
physical system. Particularly the set of singular points of Green's
function coincides with the spectrum. Green's function is in fact
the integral kernel of the resolvent of the Hamiltonian which is regarded
as an integral operator. Sometimes, however, the integral kernel should
be interpreted in the distributional sense. Green's function is also
closely related to the heat kernel or to the propagator. Namely, Green's
function is the Laplace transform of the heat kernel.

Green's function can be explicitly expressed in a compact form for
some quantum systems which are usually distinguished by their symmetry
properties. As a rule, such systems frequently enjoy rotational symmetry.
If so, this also opens the way to an alternative construction of Green's
function based on the method of separation of variables. The problem
then effectively reduces to a one-dimensional one. Finally one deals
with a positive second-order ordinary differential operator of Sturm-Liouville
type though on the half-line rather than on a bounded interval. This
is a substantial simplification since the construction of Green's
function for a Sturm-Liouville operator is commonly known and, in
fact, this is a text-book matter. Thus in distinguished solvable cases
the radial part of Green's function can be expressed in terms of appropriate
special functions. The full Hamiltonian depending on both radial and
angle variables is then expressed as a sum with the summation running
over eigenmodes of the spherical part of the Hamiltonian. Equaling
the compact form of Green's function to the sum obtained via the method
of separation of variables leads to an addition formula for the involved
special functions.

This procedure can be successfully applied to the Hamiltonian of the
hydrogen atom, resulting in an addition formula for the Whittaker
functions $M_{\kappa,\mu}$ and $W_{\kappa,\mu}$. The formula turns
out to be a generalization of Gegenbauer's Addition Theorem in some
respect. To the best of author's knowledge, this possibility has not
been exploited yet and still remains overlooked. And this is despite
the fact that a compact formula for Coulomb Green's function has been
derived by Hostler rather long time ago in \cite{Hostler,HostlerPratt}.
As a companion of the addition formula for the Whittaker functions
we further derive another addition formula concerning the Laguerre
polynomials. There already exists a well-known addition formula for
the Laguerre polynomials but that one reported here is completely
different.

From the mathematical point of view, the formula for the Whittaker
functions cannot be considered fully satisfactory, however, as the
resulting sum involves only the Whittaker functions with the parameter
$\mu=1/2$. A more general formula for arbitrary parameters $\kappa$
and $\mu$ seems to be lacking. Nevertheless here we present some
partial results in this direction which indicate that the derived
formula could be further generalized. In addition to the parameter
$\kappa$, with $\mu$ being restricted to the values $1/2$ modulo
integers, the formula depends on the radial variables $r_{0}$ and
$r$ and on an angle $\gamma$. In the particular case $\gamma=\pi$
we show that there exists an addition formula admitting general values
of both $\kappa$ and $\mu$. Further we derive a summation formula
for the Whittaker functions $W_{\kappa,\mu}$ only and another one
for the Whittaker functions $M_{\kappa,\mu}$. Again, in both cases,
the parameters $\kappa$ and $\mu$ can take arbitrary values.

The paper is organized as follows. In Section II we summarize some
known formulas and results which are essential for the solution of
our problem. The main result of the paper, namely a derivation of
an addition formula for the Whittaker functions, is the content of
Section III. Section IV is devoted to an addition formula for the
Laguerre polynomials. Finally, Section V contains some complementary
results suggesting that further generalizations could be possible,
as discussed above.

\section{{\large{}PRELIMINARIES}\label{sec:Preliminaries}}

First let us recall the definition of the Whittaker functions and
summarize several useful formulas related to them \cite{AbramowitzStegun,GradshteynRyzhik}.
The Whittaker functions are defined in terms of the confluent hypergeometric
functions, 
\begin{eqnarray}
M_{\kappa,\mu}(r) & := & e^{-r/2}\,r^{1/2+\mu}\,_{1}F_{1}\!\left(\mu-\kappa+\frac{1}{2};2\mu+1;r\right),\label{eq:M-def}\\
W_{\kappa,\mu}(r) & := & e^{-r/2}\,r^{1/2+\mu}\,U\!\left(\mu-\kappa+\frac{1}{2},2\mu+1,r\right)\!.\label{eq:W-def}
\end{eqnarray}
For the derivatives we shall use the notation
\[
W_{\nu,1/2}'(x):=\frac{\partial W_{\nu,1/2}(x)}{\partial x},\ M_{\nu,1/2}'(x):=\frac{\partial M_{\nu,1/2}(x)}{\partial x}\,.
\]
The modified Bessel functions $K_{\nu}$, $I_{\nu}$ are particular
cases of the Whittaker functions,`
\begin{equation}
K_{\nu}(z)=\sqrt{\frac{\pi}{2z}}\,W_{0,\nu}(2z),\ I_{\nu}(z)=\frac{1}{2^{2\nu}\Gamma(\nu+1)\sqrt{2z}}\,M_{0,\nu}(2z).\label{eq:K-I-to-W-M}
\end{equation}
Furthermore,

\begin{equation}
M_{0,1/2}(z)=2\sinh\!\left(\frac{z}{2}\right),\ W_{0,1/2}(z)=e^{-z/2}.\label{eq:M-W-exp}
\end{equation}
For $n\in\mathbb{Z}_{+}$, the functions $W_{n+(\alpha+1)/2,\alpha/2}$
and $M_{n+(\alpha+1)/2,\alpha/2}$ are linearly dependent,
\[
M_{n+(\alpha+1)/2,\alpha/2}(z)=\frac{(-1)^{n}\Gamma(\alpha+1)}{\Gamma(n+\alpha+1)}\,W_{n+(\alpha+1)/2,\alpha/2}(z).
\]
Moreover, the generalized (associated) Laguerre polynomials are related
to the Whittaker functions,
\[
L_{n}^{\alpha}(z)=\frac{(-1)^{n}}{n!}\,z^{-(\alpha+1)/2}W_{n+(\alpha+1)/2,\alpha/2}(z).
\]
Therefore, for $n\in\mathbf{Z}_{+}$,
\begin{equation}
W_{n,1/2}(z)=(-1)^{n+1}n!\,M_{n,1/2}(z)\label{eq:WM}
\end{equation}
and, for $n\geq1$,
\begin{equation}
M_{n,1/2}(z)=\frac{1}{n}\,e^{-z/2}zL_{n-1}^{1}(z).\label{eq:ML}
\end{equation}
Regarding the asymptotic forms, we have
\begin{equation}
e^{r/2}\,W_{\kappa,\mu}(r)=r^{\kappa}\left(1+\frac{\left(\mu-\kappa+\frac{1}{2}\right)\left(\mu+\kappa-\frac{1}{2}\right)}{r}+O\!\left(\frac{1}{r^{2}}\right)\!\right)\!,\,\ \text{as}\ r\to\infty\label{eq:asympt-W-r}
\end{equation}
(see, for example, (\ref{eq:W-def}) and equation 13.5.2 in \cite{AbramowitzStegun}).
Furthermore \cite[Eqs. 13.20.1, 13.20.2]{DLMF}, when $\mu\to\infty$
in $\mathbb{C}$, $\Re(\mu)>0$, and $\kappa\in\mathbb{C}$ is fixed,
\begin{equation}
M_{\kappa,\mu}(z)=z^{\mu+1/2}\left(1+O(\mu^{-1})\right)\label{eq:M-asympt}
\end{equation}
uniformly for $z$ in a bounded region in $\mathbb{C}$, and
\begin{equation}
W_{\kappa,\mu}(x)=\frac{\Gamma(\kappa+\mu)}{\sqrt{\pi}}\left(\frac{x}{4}\right)^{1/2-\mu}\left(1+O(\mu^{-1})\right)\label{eq:W-asympt}
\end{equation}
uniformly for bounded positive values of $x$ (one can also consult
\cite[Chap. 7, Sect. 11.1]{Olver} or \cite{Dunster-MW}).

Next let us shortly recall, without going into all details, a standard
construction of Green's function of a Sturm-Liouville operator for
it is essential for our purposes. Thus we consider a second-order
ordinary differential operator
\[
Lf(x):=-\big(p(x)f'(x)\big)'+q(x)f(x)
\]
on an interval which can be bounded or unbounded. Here $p(x)>0$ and
$q(x)\geq0$ are sufficiently regular functions. At the finite endpoints
one imposes mixed boundary conditions or, if the interval is unbounded,
one requires functions from the domain of $L$ to be square integrable
on a neighborhood of infinity. One assumes that $L$ with properly
chosen boundary conditions is positive definite. To describe Green's
function of $L$ one finds two nontrivial solutions $v_{0}$, $v_{1}$
of the differential equation
\[
-pv_{j}''-p'v_{j}'+qv_{j}=0,\ j=0,1,
\]
on the given interval such that $v_{0}$ satisfies the boundary condition
at the left endpoint (or minus infinity) only while $v_{1}$ satisfies
the boundary condition at the right endpoint (or plus infinity) only.
Then $L^{-1}$ is an integral operator with the integral kernel
\[
\mathcal{G}(x,y)=-\frac{1}{pw}\,\left(\vartheta(y-x)v_{0}(x)v_{1}(y)+\vartheta(x-y)v_{0}(y)v_{1}(x)\right)
\]
where $w:=v_{0}v_{1}'-v_{1}v_{0}'$ is the Wronskian of $v_{0}$ and
$v_{1}$. Note that $p(x)w(x)$ is in fact a constant function. Here
and in the sequel $\vartheta$ denotes the Heaviside step function.

It may be instructive to illustrate the procedure leading to an addition
formula, as described in Section I, on the well-known example of the
operator $-\nabla^{2}+k^{2}$, $k>0$, in $\mathbb{R}^{2}$. Naturally,
the partial differential operator is expressed in polar coordinates.
Using the method of separation of variables one finds that Green's
function can be written in the form
\begin{equation}
G(r,\varphi;r_{0},\varphi_{0})=\frac{1}{2\pi\sqrt{rr_{0}}}\left(f_{0}(r,r_{0})+2\sum_{n=1}^{\infty}f_{n}(r,r_{0})\cos(n(\varphi-\varphi_{0}))\right).\label{eq:G-2d-free}
\end{equation}
The functions $f_{n}(r,r_{0})$ can be obtained as solutions of the
Sturm-Liouville problem in the radial variable, as described above,
and we get
\begin{eqnarray*}
 &  & \hskip-2.5emf_{n}(r,r_{0})=\frac{1}{2k(2n)!}\,\Gamma\!\left(n+\frac{1}{2}\right)\\
\noalign{\smallskip} &  & \hskip2.5em\times\,\big(\vartheta(r-r_{0})W_{0,n}(2kr)M_{0,n}(2kr_{0})+\vartheta(r_{0}-r)M_{0,n}(2kr)W_{0,n}(2kr_{0})\big).
\end{eqnarray*}
The RHS can be also expressed in terms of the modified Bessel functions,
see (\ref{eq:K-I-to-W-M}). At the same time, a compact formula for
Green's function is well known,
\[
G(\pmb{r},\pmb{r}_{0})=\frac{1}{2\pi}\,K_{0}(k|\,\pmb{r}-\pmb{r}_{0}|).
\]
We can let $\varphi_{0}=0$ and, by comparison, we obtain an addition
formula for the modified Bessel functions,
\begin{equation}
I_{0}(kr_{0})K_{0}(kr)+2\sum_{n=1}^{\infty}I_{n}(kr_{0})K_{n}(kr)\cos(n\varphi)=K_{0}(kR)\ \,\text{for\ }0\leq r_{0}<r,\label{eq:ModBessel-add}
\end{equation}
where
\begin{equation}
R=R(\varphi):=\sqrt{r^{2}+r_{0}^{\,2}-2rr_{0}\cos(\varphi)}\,.\label{eq:R}
\end{equation}
As a matter of fact, formula (\ref{eq:ModBessel-add}) is a corollary
of substantially more general Graf's Addition Theorem \cite[Eq. 9.1.79]{AbramowitzStegun}.

Let us rewrite (\ref{eq:G-2d-free}) just to have a comparison to
Hostler's result (\ref{eq:D-Green-Hostler}) which is mentioned below.
From (\ref{eq:ModBessel-add}) we deduce that
\[
I_{0}(v)K_{0}(u)+2\sum_{n=1}^{\infty}I_{n}(v)K_{n}(u)=K_{0}(u-v)\ \ \text{for}\ 0\leq v<u.
\]
Let ($r=|\pmb{r}|$, $r_{0}=\left|\pmb{r}_{0}\right|$) 
\begin{equation}
x:=r+r_{0}+|\pmb{r}-\pmb{r}_{0}|,\ y:=r+r_{0}-|\pmb{r}-\pmb{r}_{0}|.\label{eq:H-Hostler-x-y}
\end{equation}
Then
\[
2\pi\,G(\pmb{r},\pmb{r}_{0})=I_{0}\!\left(\frac{ky}{2}\right)K_{0}\!\left(\frac{kx}{2}\right)+2\sum_{n=1}^{\infty}I_{n}\!\left(\frac{ky}{2}\right)K_{n}\!\left(\frac{kx}{2}\right)\!.
\]

One can proceed very analogously in case of the operator $-\nabla^{2}+k^{2}$,
$k>0$, in $\mathbb{R}^{3}$. Doing so one obtains a particular case
of Gegenbauer's Addition Theorem. We are not going to discuss this
case, however. Instead, in Section III, we will focus on the operator
$-\nabla^{2}-g/|\pmb{r}|+k^{2}$, $g>0$ and $k>0$, in $\mathbb{R}^{3}$.
Nevertheless let us recall what Gegenbauer's theorem claims if specialized
to the modified Bessel functions. For $0\leq r_{0}<r$ and $\nu\in\mathbb{R}$,
\begin{equation}
\frac{2^{\nu}\Gamma(\nu)}{(rr_{0})^{\nu}}\sum_{n=0}^{\infty}(\nu+n)K_{\nu+n}(r)I_{\nu+n}(r_{0})C_{n}^{(\nu)}(\cos(\gamma))=\frac{K_{\nu}(R)}{R^{\nu}}\label{eq:Gegenbauer}
\end{equation}
with $R$ defined in (\ref{eq:R}), see \cite[\S II.4]{Watson} (and
also \cite[Eq. 9.1.80]{AbramowitzStegun}). Here $C_{n}^{(\nu)}(z)$
are the Gegenbauer polynomials.

Another addition formula which is crucial for our purposes is Spherical
Harmonic Addition Theorem, also called Legendre Addition Theorem.
Recall that the spherical harmonics are defined for $\ell\in\mathbb{Z}_{+}$
(the set of non-negative integers), $m\in\mathbb{Z}$, $|m|\leq\ell$,

\begin{equation}
Y_{\ell}^{m}(\theta,\varphi):=\sqrt{\frac{(2\ell+1)(\ell-m)!}{4\pi\,(\ell+m)!}}\,P_{\ell}^{m}(\cos(\theta))e^{im\varphi}.\label{eq:C-Yml}
\end{equation}
Here $\theta\in[\,0,\pi\,]$, $\varphi\in[\,0,2\pi\,]$ are coordinates
on the unit sphere $S^{2}$ and $P_{\ell}^{m}(z)$ is the associated
Legendre polynomial. We have
\begin{equation}
P_{\ell}^{-m}(z)=(-1)^{m}\frac{(\ell-m)!}{(\ell+m)!}\,P_{\ell}^{m}(z).\label{eq:B-P-ml}
\end{equation}
and
\[
Y_{\ell}^{-m}(\theta,\varphi)=(-1)^{m}\,\overline{Y_{\ell}^{m}(\theta,\varphi)}.
\]
The spherical harmonics $\{Y_{\ell}^{m}\}$ form an orthonormal basis
in $L^{2}(S^{2},\text{d}\Omega)$ and
\[
-\nabla^{2}\,Y_{\ell}^{m}(\theta,\varphi)=\frac{\ell\,(\ell+1)}{r^{2}}\,Y_{\ell}^{m}(\theta,\varphi).
\]

Spherical Harmonic Addition Theorem tells us that, for every $\ell\in\mathbb{Z}_{+}$,
\begin{equation}
\sum_{m=-\ell}^{\ell}Y_{\ell}^{m}(\theta,\varphi)\overline{Y_{\ell}^{m}(\theta_{0},\varphi_{0})}=\sum_{m=-\ell}^{\ell}(-1)^{m}Y_{\ell}^{m}(\theta,\varphi)Y_{\ell}^{-m}(\theta_{0},\varphi_{0})=\frac{2\ell+1}{4\pi}\,P_{\ell}(\cos(\gamma))\label{eq:Y-Harmonic-add}
\end{equation}
where
\[
\cos(\gamma):=\cos(\theta)\cos(\theta_{0})+\sin(\theta)\sin(\theta_{0})\cos(\varphi-\varphi_{0}),
\]
that is $\cos(\gamma)=\pmb{n}\cdot\pmb{n}_{0}$, $\pmb{n}:=\big(\sin(\theta)\cos(\varphi),\sin(\theta)\sin(\varphi),\cos(\theta)\big)$
and $\pmb{n}_{0}$ is defined similarly. $P_{\ell}(z)\equiv P_{\ell}^{0}(z)$
are the Legendre polynomials. Referring to (\ref{eq:C-Yml}) and (\ref{eq:B-P-ml}),
equation (\ref{eq:Y-Harmonic-add}) means that
\begin{eqnarray*}
P_{\ell}(\cos(\gamma)) & = & P_{\ell}(\cos(\theta))P_{\ell}(\cos(\theta_{0}))\\
 &  & +\,2\sum_{m=1}^{\ell}\frac{(\ell-m)!}{(\ell+m)!}\,P_{\ell}^{m}(\cos(\theta))P_{\ell}^{m}(\cos(\theta_{0}))\cos(m(\varphi-\varphi_{0})).
\end{eqnarray*}

This is a classical result with a long history \cite{Ferrers}, and
a dozen different proofs of it have been provided, some of them quite
intricate \cite{MalecekNadenik}. A straightforward derivation, as
encountered in physical literature, is based on symmetry considerations
and can be briefly rephrased as follows. Observe that
\[
\tilde{\mathcal{P}}_{\ell}(\pmb{n},\pmb{n}_{0}):=\sum_{m=-\ell}^{\ell}Y_{\ell}^{m}(\theta,\varphi)\overline{Y_{\ell}^{m}(\theta_{0},\varphi_{0})}
\]
is the integral kernel of the orthogonal projection in $L^{2}(S^{2},d\Omega)$
onto the eigenspace of minus the Laplace-Beltrami operator on $S^{2}$
(denoted as $-\Delta_{S^{2}}$) corresponding to the eigenvalue $\ell(\ell+1)$.
Thus we have (the differential operator acts in variables $\theta$
and $\varphi$)
\begin{equation}
-\Delta_{S^{2}}\tilde{\mathcal{P}}_{\ell}(\pmb{n},\pmb{n}_{0})=\ell(\ell+1)\tilde{\mathcal{P}}_{\ell}(\pmb{n},\pmb{n}_{0}).\label{eq:LB-S2-eigenval}
\end{equation}
Owing to rotational symmetry $\tilde{\mathcal{P}}_{\ell}(\pmb{n},\pmb{n}_{0})$
should depend on the distance of the points $\pmb{n},\pmb{n}_{0}\in S^{2}$
only which in turn is a function of $\pmb{n}\cdot\pmb{n}_{0}=\cos(\gamma)$.
Writing $\tilde{\mathcal{P}}_{\ell}(\pmb{n},\pmb{n}_{0})=f(\pmb{n}\cdot\pmb{n}_{0})$
equation (\ref{eq:LB-S2-eigenval}) reduces to the ordinary second-order
differential equation
\[
-(1-z^{2})f''(z)+2zf'(z)-\ell\,(\ell+1)f(z)=0.
\]
A general solution has the form $f(z)=c_{1}P_{\ell}(z)+c_{2}Q_{\ell}(z)$.
Here $Q_{\ell}(z)$ is the Legendre function of the second kind which
is another independent solutions of the differential equation and
is known to be singular for $z=\pm1$. Therefore $\tilde{\mathcal{P}}_{\ell}(\pmb{n},\pmb{n}_{0})=c_{1}P_{\ell}(\pmb{n}\cdot\pmb{n}_{0})$.
The multiplicative constant is easily found to be $c_{1}=(2\ell+1)/(4\pi)$
when letting $\pmb{n}=\pmb{n}_{0}=(0,0,1)$ and taking into account
that $P_{\ell}(1)=1$ and $P_{\ell}^{m}(1)=0$ for $m\neq0$.

\section{{\large{}THE HYDROGEN ATOM AND AN ADDITION\newline FORMULA FOR THE
WHITTAKER FUNCTIONS}\label{sec:The-hydrogen-atom}}

Using spherical coordinates $r>0$, $\theta\in[\,0,\pi\,]$, $\varphi\in[\,0,2\pi\,]$,
we denote
\[
\pmb{r}=\big(r\sin(\theta)\cos(\varphi),r\sin(\theta)\sin(\varphi),r\cos(\varphi)\big).
\]
We write the Hamiltonian of the hydrogen atom in a dimensionless form
as $H=-\nabla^{2}-g/r$, $g>0$, and we wish to apply the procedure
leading to an addition formula, as described in Section I, to the
operator
\[
H+k^{2}=-\frac{1}{r^{2}}\frac{\partial}{\partial r}r^{2}\frac{\partial}{\partial r}-\frac{1}{r^{2}}\!\left(\frac{1}{\sin(\theta)}\frac{\partial}{\partial\theta}\sin(\theta)\frac{\partial}{\partial\theta}+\frac{1}{\sin^{2}(\theta)}\frac{\partial^{2}}{\partial\varphi^{2}}\right)\!-\frac{g}{r}+k^{2},\ \,k>0.
\]
While the continuous spectrum of $H$ coincides with the positive
real half-line, the discrete spectrum consists of eigenvalues $E_{n}=-g^{2}/4n^{2}$,
$n\in\mathbb{N}$, the multiplicity of $E_{n}$ equals $n^{2}$. The
corresponding normalized eigenfunctions are
\[
\psi_{n,\ell,m}(r,\theta,\varphi)=\frac{g^{3/2}}{n^{\ell+2}}\,\sqrt{\frac{(n-\ell-1)!}{2(n+\ell)!}}\,(gr)^{\ell}\exp\!\left(-\frac{gr}{2n}\right)L_{n-\ell-1}^{2\ell+1}\!\left(\frac{gr}{n}\right)Y_{\ell}^{m}(\theta,\varphi),
\]
where $n\in\mathbb{N}$, $\ell\in\mathbb{Z}_{+}$ and $m\in\mathbb{Z}$
are the principal, the azimuthal and the magnetic quantum number,
respectively, and $|m|\leq\ell\leq n-1$.

Application of the method of separation of variables to this operator
again leads to the Sturm-Liouville problem in the radial variable
whose solution is rather straightforward, as outlined in Section II,
and has already been described in the literature \cite{VasanSeetharaman,Rauh}.
We have
\begin{eqnarray*}
 &  & \hskip-2.3emG(r,\theta,\varphi,r_{0},\theta_{0},\varphi_{0})=\frac{1}{rr_{0}}\sum_{\ell=0}^{\infty}\sum_{m=-\ell}^{\ell}\frac{1}{2k\,(2\ell+1)!}\,\Gamma\!\left(\ell+1-\frac{g}{2k}\right)\\
\noalign{\smallskip} &  & \hskip3em\times\big(\vartheta(r-r_{0})M_{g/(2k),\ell+1/2}(2kr_{0})W_{g/(2k),\ell+1/2}(2kr)\\
\noalign{\smallskip} &  & \hskip3em\quad\ +\,\vartheta(r_{0}-r)M_{g/(2k),\ell+1/2}(2kr)W_{g/(2k),\ell+1/2}(2kr_{0})\big)\,Y_{\ell}^{m}(\theta,\varphi)\overline{Y_{\ell}^{m}(\theta_{0},\varphi_{0})}\,.
\end{eqnarray*}
With the aid of (\ref{eq:Y-Harmonic-add}) this equation can be further
simplified,
\begin{eqnarray}
 &  & \hskip-3.2emG(r,\theta,\varphi,r_{0},\theta_{0},\varphi_{0})=\frac{1}{8\pi krr_{0}}\sum_{\ell=0}^{\infty}\frac{1}{(2\ell)!}\,\Gamma\!\left(\ell+1-\frac{g}{2k}\right)\nonumber \\
\noalign{\smallskip} &  & \hskip6em\times\big(\vartheta(r-r_{0})M_{g/(2k),\ell+1/2}(2kr_{0})W_{g/(2k),\ell+1/2}(2kr)\label{eq:G-hydrogen-separation}\\
\noalign{\smallskip} &  & \hskip6em\quad\ +\,\vartheta(r_{0}-r)M_{g/(2k),\ell+1/2}(2kr)W_{g/(2k),\ell+1/2}(2kr_{0})\big)P_{\ell}(\cos(\gamma)).\nonumber 
\end{eqnarray}

Notably, there exists a remarkable compact formula for Green's function
due to Hostler \cite{Hostler,HostlerPratt},
\begin{eqnarray}
 &  & \hskip-3emG(r,\theta,\varphi,r_{0},\theta_{0},\varphi_{0})=\frac{1}{4\pi R}\,\Gamma\!\left(1-\frac{g}{2k}\right)\label{eq:D-Green-Hostler}\\
\noalign{\medskip} &  & \hskip4em\times\left(M_{g/(2k),1/2}'(ky)W_{g/(2k),1/2}(kx)-M_{g/(2k),1/2}(ky)W_{g/(2k),1/2}'(kx)\right)\nonumber 
\end{eqnarray}
where in this case we have
\[
R:=|\pmb{r}-\pmb{r}_{0}|=\sqrt{r^{2}+r_{0}^{\,2}-2rr_{0}\big(\sin(\theta)\sin(\theta_{0})\cos(\varphi-\varphi_{0})+\cos(\theta)\cos(\theta_{0})\big)}\,,
\]
and
\begin{equation}
x:=r+r_{0}+|\pmb{r}-\pmb{r}_{0}|,\ y:=r+r_{0}-|\pmb{r}-\pmb{r}_{0}|\label{eq:H-Hostler-x-y-1}
\end{equation}
(formally the same equations as in (\ref{eq:H-Hostler-x-y}) but now
the dimension is $3$ rather than $2$).

Comparing (\ref{eq:G-hydrogen-separation}) to (\ref{eq:D-Green-Hostler})
while still using notation (\ref{eq:H-Hostler-x-y-1}) we have, for
$0\leq r_{0}<r$,
\begin{eqnarray*}
 &  & \frac{1}{krr_{0}}\sum_{\ell=0}^{\infty}\frac{1}{(2\ell)!}\,\Gamma\!\left(\ell+1-\frac{g}{2k}\right)M_{g/(2k),\ell+1/2}(2kr_{0})W_{g/(2k),\ell+1/2}(2kr)P_{\ell}(\cos(\gamma))\\
 &  & =\Gamma\!\left(1-\frac{g}{2k}\right)\frac{2}{R}\,\big(M_{g/(2k),1/2}'(ky)W_{g/(2k),1/2}(kx)-M_{g/(2k),1/2}(ky)W_{g/(2k),1/2}'(kx)\big).
\end{eqnarray*}
After substitution $g=2k\kappa$ and rescaling $r\to r/(2k)$, $r_{0}\to r_{0}/(2k)$
we get an addition formula for the Whittaker functions. Moreover,
the parameter $\kappa$ can be extended to complex values by analyticity.

\begin{theorem}\label{thm:Whittaker-Add} For $0\leq r_{0}<r$, $\kappa\in\mathbb{C}\setminus\mathbb{N}$
and $\gamma\in\mathbb{R}$,
\begin{eqnarray}
 &  & \hskip-1em\frac{1}{rr_{0}}\sum_{\ell=0}^{\infty}\frac{\Gamma(\ell+1-\kappa)}{\Gamma(1-\kappa)(2\ell)!}\,M_{\kappa,\ell+1/2}(r_{0})W_{\kappa,\ell+1/2}(r)P_{\ell}(\cos(\gamma))\nonumber \\
 &  & \hskip-1em=\frac{1}{R}\left(M_{\kappa,1/2}'\!\left(\frac{y}{2}\right)W_{\kappa,1/2}\!\left(\frac{x}{2}\right)-M_{\kappa,1/2}\!\left(\frac{y}{2}\right)W_{\kappa,1/2}'\!\left(\frac{x}{2}\right)\right)\label{eq:Whittaker-add}
\end{eqnarray}
where $R=R(\gamma)$, see (\ref{eq:R}), and
\begin{equation}
x=r+r_{0}+R,\ y=r+r_{0}-R.\label{eq:H-Hostler-x-y-2}
\end{equation}
\end{theorem}

One has to exclude the values $\kappa\in\mathbb{N}$. As a matter
of fact, the equation holds for these values, too, but it should be
achieved in the limit after the singular terms in the equation have
been combined. For instance, for $\kappa=1$ we have
\begin{eqnarray*}
 &  & \frac{1}{rr_{0}}\sum_{\ell=1}^{\infty}\frac{(\ell-1)!}{(2\ell)!}\,M_{1,\ell+1/2}(r_{0})W_{1,\ell+1/2}(r)P_{\ell}\big(\cos(\gamma)\big)\\
 &  & =-\frac{\partial}{\partial\kappa}\Bigg(\frac{1}{R}\bigg(M_{\kappa,1/2}'\left(\frac{y}{2}\right)W_{\kappa,1/2}\!\left(\frac{x}{2}\right)-M_{\kappa,1/2}\!\left(\frac{y}{2}\right)W_{\kappa,1/2}'\!\left(\frac{x}{2}\right)\bigg)\\
 &  & \qquad\qquad\quad-\frac{1}{rr_{0}}\,M_{\kappa,1/2}(r_{0})W_{\kappa,1/2}(r)\Bigg)\bigg|_{\kappa=1}.
\end{eqnarray*}

\begin{remark} Let us check two particular cases. For $\gamma=0$
(hence $R=r-r_{0}$, $x=2r$, $y=2r_{0}$) we have
\begin{eqnarray}
 &  & \frac{1}{rr_{0}}\sum_{\ell=0}^{\infty}\frac{\Gamma(\ell+1-\kappa)}{(2\ell)!}\,M_{\kappa,\ell+1/2}(r_{0})W_{\kappa,\ell+1/2}(r)\nonumber \\
 &  & =\frac{\Gamma(1-\kappa)}{r-r_{0}}\,\big(M_{\kappa,1/2}'(r_{0})W_{\kappa,1/2}(r)-M_{\kappa,1/2}(r_{0})W_{\kappa,1/2}'(r)\big),\label{eq:Whittaker-part-gamma-0}
\end{eqnarray}
and for $\gamma=\pi$ (hence $R=r+r_{0}$, $x=2(r+r_{0})$, $y=0$)
we have
\begin{equation}
\frac{1}{rr_{0}}\sum_{\ell=0}^{\infty}(-1)^{\ell}\,\frac{\Gamma(\ell+1-\kappa)}{(2\ell)!}\,M_{\kappa,\ell+1/2}(r_{0})W_{\kappa,\ell+1/2}(r)=\frac{\Gamma(1-\kappa)}{r+r_{0}}\,W_{\kappa,1/2}(r+r_{0}).\label{eq:Whittaker-part-gamma-pi}
\end{equation}
\end{remark}

\begin{remark} From (\ref{eq:Whittaker-part-gamma-pi}) one can derive
a summation formula for Whittaker functions which seems to be also
new. For $\kappa\in\mathbb{C}\setminus\mathbb{N}$ and $z\in\mathbb{C}$,
\begin{equation}
\frac{1}{z}\sum_{\ell=0}^{\infty}(-1)^{\ell}\,\frac{\Gamma(\ell+1-\kappa)}{\Gamma(1-\kappa)(2\ell)!}\,M_{\kappa,\ell+1/2}(z)=e^{-z/2}\label{eq:E-summ-M}
\end{equation}
(where the singularity at $z=0$ on the LHS is removable). It can
be proven by exploring the asymptotic behavior of both sides of (\ref{eq:Whittaker-part-gamma-pi}),
as $r\to\infty$. We have, in virtue of (\ref{eq:asympt-W-r}),
\[
\frac{1}{r+r_{0}}\,W_{\kappa,1/2}(r+r_{0})=e^{-r/2-r_{0}/2}r^{-1+\kappa}\left(1+\frac{\kappa-\kappa^{2}-(1-\kappa)r_{0}}{r}+O\!\left(\frac{1}{r^{2}}\right)\right)
\]
and
\[
\frac{1}{r}\,W_{\kappa,\ell+1/2}(r)=e^{-r/2}r^{-1+\kappa}\left(1+\frac{(\ell+1-\kappa)(\ell+\kappa)}{r}+O\!\left(\frac{1}{r^{2}}\right)\right)\!.
\]
Thus, writing $z$ instead of $r_{0}$, we see that (\ref{eq:E-summ-M})
holds for $z>0$. But the asymptotic behavior of $M_{\kappa,\ell+1/2}(z)$
for $\ell$ large, as recalled in (\ref{eq:M-asympt}), which is locally
uniform in $z$ implies that the LHS is an entire function of $z$.
Since the same is true for the RHS we conclude that (\ref{eq:E-summ-M})
must hold for all $z\in\mathbb{C}$. \end{remark}

Let us point out a relation of Theorem \ref{thm:Whittaker-Add} to
Gegenbauer's Addition Theorem. Confining ourselves to the value $\kappa=0$
(corresponding to $g=0$) in (\ref{eq:Whittaker-add}) we obtain a
simplified equation
\begin{equation}
\sum_{\ell=0}^{\infty}\frac{2\ell+1}{\sqrt{rr_{0}}}\,K_{\ell+1/2}\!\left(\frac{r}{2}\right)I_{\ell+1/2}\!\left(\frac{r_{0}}{2}\right)P_{\ell}(\cos(\gamma))=\frac{1}{R}\,e^{-R/2}=\frac{1}{\sqrt{\pi R}}\,K_{1/2}\!\left(\frac{R}{2}\right)\!.\label{eq:Gehenbauer-Add-Thm-part}
\end{equation}
This is a particular case of Gegenbauer's Addition Theorem, however,
see equation (\ref{eq:Gegenbauer}) with $\nu=1/2$. To derive (\ref{eq:Gehenbauer-Add-Thm-part})
from (\ref{eq:Whittaker-add}) we have used (\ref{eq:K-I-to-W-M}),
(\ref{eq:M-W-exp}) and also the equation
\[
M_{0,1/2}'\!\left(\frac{y}{2}\right)W_{0,1/2}\!\left(\frac{x}{2}\right)-M_{0,1/2}\!\left(\frac{y}{2}\right)W_{0,1/2}'\!\left(\frac{x}{2}\right)=e^{-R/2}=\sqrt{\frac{R}{\pi}}\,K_{1/2}\!\left(\frac{R}{2}\right)\!.
\]
Moreover, note that $C_{n}^{(1/2)}(z)=P_{n}(z)$.

As for equation (\ref{eq:E-summ-M}), letting $\kappa=0$ we get
\[
\sqrt{\frac{\pi}{2r}}\sum_{\ell=0}^{\infty}(-1)^{\ell}(2\ell+1)I_{\ell+1/2}(z)=e^{-z}.
\]
This is a particular case of the identity
\[
2^{\nu}\Gamma(\nu)\sum_{\ell=0}^{\infty}(\ell+\nu)C_{\ell}^{(\nu)}(\gamma)I_{\ell+\nu}(z)=z^{\nu}e^{\gamma z},
\]
which is well known, see \cite[§7.15(1)]{ErdelyiMagnusOberhettingerTricomi};
note that $C_{\ell}^{(1/2)}(-1)=(-1)^{\ell}$.

\section{{\large{}AN ADDITION FORMULA FOR THE LAGUERRE\newline POLYNOMIALS}\label{sec:Addition-Laguerre}}

\begin{theorem} For $n\in\mathbb{N}$, $0\leq r_{0}<r$ and $\gamma\in\mathbb{R}$,
\begin{eqnarray}
 &  & \sum_{\ell=0}^{n-1}\frac{(2\ell+1)(n-\ell-1)!}{(n+\ell)!}\,(rr_{0})^{\ell}\,L_{n-\ell-1}^{2\ell+1}(r)L_{n-\ell-1}^{2\ell+1}(r_{0})P_{\ell}(\cos(\gamma))\nonumber \\
 &  & =\frac{1}{2R}\left(xL_{n-1}^{1}\!\left(\frac{x}{2}\right)L_{n}\!\left(\frac{y}{2}\right)-yL_{n-1}^{1}\!\left(\frac{y}{2}\right)L_{n}\!\left(\frac{x}{2}\right)\right)\label{eq:L-Laguerre-add}
\end{eqnarray}
where $x$, $y$ are defined in (\ref{eq:H-Hostler-x-y-2}), with
$R=R(\gamma)$, see (\ref{eq:R}). \end{theorem}

\begin{remark*} Note that formula (\ref{eq:L-Laguerre-add}) is completely
different from the well known addition formula for the Laguerre polynomials
which claims that \cite[Eq. 22.12.6]{AbramowitzStegun}
\[
\sum_{j=0}^{n}L_{j}^{\alpha}(u)L_{n-j}^{\beta}(v)=L_{n}^{\alpha+\beta+1}(u+v).
\]
\end{remark*}

\begin{proof} Recall that
\[
\mathop{{\rm Res}}\big(\Gamma(z);\,z=-n\big)=\frac{(-1)^{n}}{n!},\ n\in\mathbb{Z}_{+},
\]
whence
\[
\mathop{{\rm Res}}\big(\Gamma\left(1-g/(2\sqrt{-z})\big);\,z=E_{n}\right)=\frac{(-1)^{n}g^{2}}{2\,n!\,n^{2}},\ n\in\mathbb{N}.
\]
If we substitute $k=\sqrt{-z}$ in Hostler's formula (\ref{eq:D-Green-Hostler})
then the residue of Green's function at $z=E_{n}$ equals
\begin{equation}
\frac{(-1)^{n}g^{2}}{8\pi n!\,n^{2}R}\left(M_{n,1/2}'\!\left(\frac{gy}{2n}\right)W_{n,1/2}\!\left(\frac{gx}{2n}\right)-M_{n,1/2}\!\left(\frac{gy}{2n}\right)W_{n,1/2}'\!\left(\frac{gx}{2n}\right)\right)\!.\label{eq:R-residuum-Green}
\end{equation}
The residue also equals minus the projection $\mathcal{P}_{n}$ onto
the eigenspace corresponding to eigenvalue $E_{n}$. $\mathcal{P}_{n}$
is an integral operator with the integral kernel
\begin{equation}
\mathcal{P}_{n}(r,\theta,\varphi,r_{0},\theta_{0},\varphi_{0})=\sum_{\ell=0}^{n-1}\sum_{m=-\ell}^{\ell}\psi_{n,\ell,m}(r,\theta,\varphi)\,\overline{\psi_{n,\ell,m}(r_{0},\theta_{0},\varphi_{0})}.\label{eq:Pn}
\end{equation}
Hence, for $n\in\mathbb{N}$,
\begin{eqnarray}
 &  & \frac{(-1)^{n+1}n}{(n-1)!\,gR}\left(M_{n,1/2}'\!\left(\frac{gy}{2n}\right)W_{n,1/2}\!\left(\frac{gx}{2n}\right)-M_{n,1/2}\!\left(\frac{gy}{2n}\right)W_{n,1/2}'\!\left(\frac{gx}{2n}\right)\right)\nonumber \\
 &  & =\exp\!\left(-\frac{g\,(r+r_{0})}{2n}\right)\sum_{\ell=0}^{n-1}\frac{(2\ell+1)(n-\ell-1)!}{(n+\ell)!}\left(\frac{g^{2}rr_{0}}{n^{2}}\right)^{\!\ell}\label{eq:Laguerre-add-pre}\\
 &  & \ \ \times\,L_{n-\ell-1}^{2\ell+1}\!\left(\frac{gr}{n}\right)L_{n-\ell-1}^{2\ell+1}\!\left(\frac{gr_{0}}{n}\right)P_{\ell}(\cos(\gamma)).\nonumber 
\end{eqnarray}
In regard of (\ref{eq:WM}) and (\ref{eq:ML}), we derive
\begin{eqnarray*}
M_{n,1/2}'(z) & = & \frac{{\rm d}}{{\rm d}z}\!\left(\frac{1}{n}\,e^{z/2}e^{-z}zL_{n-1}^{1}(z)\right)\,=\,\frac{1}{2}\,M_{n,1/2}(z)+\frac{1}{n}\,e^{z/2}\,\frac{{\rm d}}{{\rm d}z}\!\left(e^{-z}zL_{n-1}^{1}(z)\right)\\
 & = & \frac{1}{2}\,M_{n,1/2}(z)+e^{-z/2}L_{n}(z).
\end{eqnarray*}
Here we have used the formula \cite[Eq. (9.12.8)]{KoekoekLeskySwarttouw}
\[
\frac{{\rm d}}{{\rm d}z}\!\left(e^{-z}z^{\alpha}L_{n-1}^{\alpha}(z)\right)=ne^{-z}z^{\alpha-1}L_{n}^{\alpha-1}(z).
\]
Furthermore,
\begin{eqnarray}
 &  & \frac{(-1)^{n+1}g^{2}}{8\pi R\,n!\,n^{2}}\left(M_{n,1/2}'\!\left(\frac{gy}{2n}\right)W_{n,1/2}\!\left(\frac{gx}{2n}\right)-M_{n,1/2}\!\left(\frac{gy}{2n}\right)W_{n,1/2}'\!\left(\frac{gx}{2n}\right)\right)\nonumber \\
 &  & =\frac{g^{2}}{8\pi Rn^{2}}\left(M_{n,1/2}\!\left(\frac{gx}{2n}\right)M_{n,1/2}'\!\left(\frac{gy}{2n}\right)-M_{n,1/2}\!\left(\frac{gy}{2n}\right)M_{n,1/2}'\!\left(\frac{gx}{2n}\right)\right)\label{eq:Laguerre-add-LHS}\\
 &  & =\frac{g^{3}}{16\pi Rn^{4}}\exp\!\left(-\frac{g}{2n}(r+r_{0})\right)\left(xL_{n-1}^{1}\!\left(\frac{gx}{2n}\right)L_{n}\!\left(\frac{gy}{2n}\right)-yL_{n-1}^{1}\!\left(\frac{gy}{2n}\right)L_{n}\!\left(\frac{gx}{2n}\right)\right).\nonumber 
\end{eqnarray}
After rescaling $r\to(n/g)r$, $r_{0}\to(n/g)r_{0}$, we get from
(\ref{eq:Laguerre-add-pre}) and (\ref{eq:Laguerre-add-LHS})
\begin{eqnarray*}
 &  & \sum_{\ell=0}^{n-1}\frac{(2\ell+1)(n-\ell-1)!}{(n+\ell)!}\,(rr_{0})^{\ell}\,L_{n-\ell-1}^{2\ell+1}(r)L_{n-\ell-1}^{2\ell+1}(r_{0})P_{\ell}(\cos(\gamma))\\
 &  & =\frac{(-1)^{n+1}}{(n-1)!R}\exp\!\left(\frac{r+r_{0}}{2}\right)\left(M_{n,1/2}'\!\left(\frac{y}{2}\right)W_{n,1/2}\!\left(\frac{x}{2}\right)-M_{n,1/2}\!\left(\frac{y}{2}\right)W_{n,1/2}'\!\left(\frac{x}{2}\right)\right)\\
 &  & =\frac{1}{2R}\left(xL_{n-1}^{1}\!\left(\frac{x}{2}\right)L_{n}\!\left(\frac{y}{2}\right)-yL_{n-1}^{1}\!\left(\frac{y}{2}\right)L_{n}\!\left(\frac{x}{2}\right)\right).
\end{eqnarray*}
This concludes the proof. \end{proof}

Let us check two particular cases of (\ref{eq:L-Laguerre-add}). For
$\gamma=0$ we have
\begin{eqnarray*}
 &  & \sum_{\ell=0}^{n-1}\frac{(2\ell+1)(n-\ell-1)!}{(n+\ell)!}\,(rr_{0})^{\ell}\,L_{n-\ell-1}^{2\ell+1}(r)L_{n-\ell-1}^{2\ell+1}(r_{0})\\
 &  & =\frac{1}{r-r_{0}}\big(rL_{n-1}^{1}(r)L_{n}(r_{0})-r_{0}L_{n-1}^{1}(r_{0})L_{n}(r)\big)
\end{eqnarray*}
and for $\gamma=\pi$ we get
\[
\sum_{\ell=0}^{n-1}(-1)^{\ell}\,\frac{(2\ell+1)(n-\ell-1)!}{(n+\ell)!}\,(rr_{0})^{\ell}\,L_{n-\ell-1}^{2\ell+1}(r)L_{n-\ell-1}^{2\ell+1}(r_{0})=L_{n-1}^{1}(r+r_{0}),
\]
Note that $L_{n}(0)=1$ (and $L_{n}^{1}(0)=n+1$).

After the shift $n\to n+1$ one observes, in these two particular
cases, that both sides are symmetric polynomials in $r$ and $r_{0}$,
and therefore $r$, $r_{0}$ can be replaced by arbitrary complex
variables.

\begin{corollary} For every $n\in\mathbb{Z}_{+}$ and all $u,v\in\mathbb{C}$,
\begin{equation}
\sum_{\ell=0}^{n}\frac{(2\ell+1)(n-\ell)!}{(n+\ell+1)!}\,(uv)^{\ell}\,L_{n-\ell}^{2\ell+1}(u)L_{n-\ell}^{2\ell+1}(v)=\frac{1}{u-v}\big(uL_{n-1}^{1}(u)L_{n}(v)-vL_{n-1}^{1}(v)L_{n}(u)\big)\label{eq:L-Laguerre-part-gamma-0}
\end{equation}
and
\begin{equation}
\sum_{\ell=0}^{n}(-1)^{\ell}\,\frac{(2\ell+1)(n-\ell)!}{(n+\ell+1)!}\,(uv)^{\ell}\,L_{n-\ell}^{2\ell+1}(u)L_{n-\ell}^{2\ell+1}(v)=L_{n}^{1}(u+v).\label{eq:L-Laguerre-part-gamma-pi}
\end{equation}
\end{corollary}

As a short digression let us note that the projection $\mathcal{P}_{n}$,
as introduced in equation (\ref{eq:Pn}) in the proof, has already
been discussed in the literature \cite{Blinder-density}. Using expression
for the residue of Green's function, see (\ref{eq:R-residuum-Green})
and (\ref{eq:Laguerre-add-LHS}), we get, for $n\geq1$,
\begin{eqnarray*}
\mathcal{P}_{n}(r,\theta,\varphi,r_{0},\theta_{0},\varphi_{0}) & = & \frac{g^{3}}{16\pi Rn^{4}}\exp\!\left(-\frac{g}{2n}(r+r_{0})\right)\\
 &  & \times\,\left(xL_{n-1}^{1}\!\left(\frac{gx}{2n}\right)L_{n}\!\left(\frac{gy}{2n}\right)-yL_{n-1}^{1}\!\left(\frac{gy}{2n}\right)L_{n}\!\left(\frac{gx}{2n}\right)\right).
\end{eqnarray*}
The diagonal in fact does not depend on angles and equals
\begin{eqnarray*}
\mathcal{P}_{n}(r,\theta,\varphi,r,\theta,\varphi) & = & \frac{g^{3}}{8\pi n^{4}}\,\exp\!\left(-\frac{gr}{n}\right)\\
 &  & \hskip-1em\times\,\left(L_{n}\!\left(\frac{gr}{n}\right)L_{n-1}^{1}\!\left(\frac{gr}{n}\right)-\frac{gr}{n}\,L_{n}\!\left(\frac{gr}{n}\right)L_{n-2}^{2}\!\left(\frac{gr}{n}\right)+\frac{gr}{n}\,L_{n-1}^{1}\!\left(\frac{gr}{n}\right){}^{2}\right)\!.
\end{eqnarray*}
$\mathcal{P}_{n}(r,\theta,\varphi,r_{0},\theta_{0},\varphi_{0})$
is called the density function in \cite{Blinder-density}, and
\[
D_{n}(r):=4\pi r^{2}\mathcal{P}_{n}(r,\theta,\varphi,r,\theta,\varphi)
\]
is called the radial distribution function. It holds true that
\[
4\pi\int_{0}^{\infty}\mathcal{P}_{n}(r,\theta,\varphi,r,\theta,\varphi)\,r^{2}{\rm d}r=n^{2}
\]
meaning that
\[
\int_{0}^{\infty}\exp(-r)\big(L_{n}(r)L_{n-1}^{1}(r)-rL_{n}(r)L_{n-2}^{2}(r)+rL_{n-1}^{1}(r){}^{2}\big)r^{2}dr=2n^{3}.
\]

\section{{\large{}SOME COMPLEMENTARY SUMMATION FORMULAS}\label{sec:Some-complementary}}

The following proposition presents a summation formula for the Whittaker
functions $W_{\kappa,\mu}$ and is a straightforward corollary of
Theorem~\ref{thm:M-Whittaker-Pi-gen} below which in turn generalizes
the addition formula (\ref{eq:Whittaker-part-gamma-pi}). Nevertheless
this proposition should be proven independently because, conversely,
it is used in the proof of Theorem~\ref{thm:M-Whittaker-Pi-gen}.

\begin{proposition}\label{thm:W-summ-W} For every $n\in\mathbb{N}$,
$r>0$, $\kappa,\mu\in\mathbb{C}$, $2\mu\neq-1,-2,-3,\ldots$,

\begin{equation}
\sum_{\ell=0}^{n}(-1)^{\ell}\binom{n}{\ell}\frac{2\mu+2\ell}{(2\mu+\ell)_{n+1}}\,W_{\kappa,\mu+\ell}(r)=(-1)^{n}r^{-n/2}\,W_{\kappa-n/2,\mu+n/2}(r).\label{eq:W-summ-W}
\end{equation}
\end{proposition}

\begin{proof} We shall proceed by induction in $n$. For $n=0$ the
equation is trivial. For $n=1$ this is a well known identity (for
instance, this is a combination of equations 9.234 ad(1) and ad(2)
in \cite{GradshteynRyzhik})
\begin{equation}
W_{\kappa,\mu+1}(r)-W_{\kappa,\mu}(r)=\frac{2\mu+1}{\sqrt{r}}\,W_{\kappa-1/2,\mu+1/2}(r).\label{eq:W1-rec}
\end{equation}
Suppose $n>0$. Set, for $k=0,1,\ldots,n$,
\begin{eqnarray*}
A(k) & := & \sum_{\ell=0}^{n-k-1}(-1)^{\ell}\binom{n}{\ell}\frac{2\mu+2\ell}{(2\mu+\ell)_{n+1}}\,W_{\kappa,\mu+\ell}(r)+(-1)^{n+k}\binom{n-1}{k}\frac{W_{\kappa,\mu+n-k}(r)}{(2\mu+n-k)_{n}}\\
 &  & -\frac{1}{\sqrt{r}}\sum_{\ell=n-k}^{n-1}(-1)^{\ell}\binom{n-1}{\ell}\frac{2\mu+2\ell+1}{(2\mu+\ell+1)_{n}}\,W_{\kappa-1/2,\mu+\ell+1/2}(r).
\end{eqnarray*}
In particular,
\begin{eqnarray*}
A(0) & = & \sum_{\ell=0}^{n-1}(-1)^{\ell}\binom{n}{\ell}\frac{2\mu+2\ell}{(2\mu+\ell)_{n+1}}\,W_{\kappa,\mu+\ell}(r)+(-1)^{n}\,\frac{W_{\kappa,\mu+n}(r)}{(2\mu+n)_{n}}\\
 & = & \sum_{\ell=0}^{n}(-1)^{\ell}\binom{n}{\ell}\frac{2\mu+2\ell}{(2\mu+\ell)_{n+1}}\,W_{\kappa,\mu+\ell}(r).
\end{eqnarray*}
Thus $A(0)$ coincides with the LHS of equation (\ref{eq:W-summ-W}).

We claim that $A(k+1)=A(k)$ for $k=0,1,\ldots,n-1$. Suppose $0\leq k<n$.
Then
\begin{eqnarray*}
A(k+1)-A(k) & = & (-1)^{n+k}\binom{n}{k+1}\frac{2\mu+2n-2k-2}{(2\mu+n-k-1)_{n+1}}\,W_{\kappa,\mu+n-k-1}(r)\\
 &  & +\,(-1)^{n+k+1}\binom{n-1}{k+1}\frac{W_{\kappa,\mu+n-k-1}(r)}{(2\mu+n-k-1)_{n}}\\
 &  & +\,(-1)^{n+k+1}\binom{n-1}{k}\frac{W_{\kappa,\mu+n-k}(r)}{(2\mu+n-k)_{n}}\\
 &  & +\,\frac{1}{\sqrt{r}}(-1)^{n+k}\binom{n-1}{k}\frac{2\mu+2n-2k-1}{(2\mu+n-k)_{n}}\,W_{\kappa-1/2,\mu+n-k-1/2}(r)\,.
\end{eqnarray*}
With the aid of (\ref{eq:W1-rec}) one finds that this expression
equals
\begin{eqnarray*}
 &  & (-1)^{n+k+1}\binom{n-1}{k+1}\frac{(2\mu+2n-k-1)W_{\kappa,\mu+n-k-1}(r)}{(2\mu+n-k-1)_{n+1}}\\
 &  & -\,(-1)^{n+k+1}\binom{n}{k+1}\frac{(2\mu+2n-2k-2)W_{\kappa,\mu+n-k-1}(r)}{(2\mu+n-k-1)_{n+1}}\\
 &  & +\,(-1)^{n+k+1}\binom{n-1}{k}\frac{(2\mu+n-k-1)W_{\kappa,\mu+n-k-1}(r)}{(2\mu+n-k-1)_{n+1}}\,.
\end{eqnarray*}
Now it is elementary to see that the expression actually equals $0$,
and therefore $\linebreak$ $A(k+1)-A(k)=0$.

In $A(n)$ we apply substitutions $\kappa'=\kappa-1/2$, $\mu'=\mu+1/2$,
and obtain, by the induction hypothesis, 
\begin{eqnarray*}
A(n) & = & -\frac{1}{\sqrt{r}}\sum_{\ell=0}^{n-1}(-1)^{\ell}\binom{n-1}{\ell}\frac{2\mu'+2\ell}{(2\mu'+\ell)_{n}}\,W_{\kappa',\mu'+\ell}(r)\\
 & = & -\frac{1}{\sqrt{r}}\,(-1)^{n-1}\,r^{-(n-1)/2}\,W_{\kappa'-(n-1)/2,\mu'+(n-1)/2}(r)\\
 & = & (-1)^{n}r^{-n/2}\,W_{\kappa-n/2,\mu+n/2}(r).
\end{eqnarray*}
Since $A(0)=A(n)$, the identity follows. \end{proof}

\begin{remark} From (\ref{eq:W-summ-W}) it follows that, for $n\in\mathbb{Z}_{+}$,
\[
\sum_{\ell=0}^{n}(-1)^{\ell}\binom{n}{\ell}\frac{2\mu+2\ell}{(2\mu+\ell)_{n+1}}=\delta_{n,0}.
\]

Indeed, one just has to recall (\ref{eq:asympt-W-r}) and compare
the asymptotic expansions of both sides of (\ref{eq:W-summ-W}) as
$r\to\infty$. \end{remark}

Here is a generalization of the addition formula for the Whittaker
functions (\ref{eq:Whittaker-part-gamma-pi}).

\begin{theorem}\label{thm:M-Whittaker-Pi-gen} For $0\leq r_{0}<r$,
$\kappa,\mu\in\mathbb{C}$, $\Re\mu>0$,
\begin{equation}
\frac{1}{(rr_{0})^{\mu+1/2}}\sum_{\ell=0}^{\infty}(-1)^{\ell}\,\frac{(\mu-\kappa+1/2)_{\ell}}{(\ell+2\mu)_{\ell}\,\ell!}\,M_{\kappa,\ell+\mu}(r_{0})W_{\kappa,\ell+\mu}(r)=\frac{1}{(r+r_{0})^{\mu+1/2}}\,W_{\kappa,\mu}(r+r_{0}).\label{eq:Whittaker-Pi-gen}
\end{equation}
\end{theorem}

\begin{remark*} The equation for $r_{0}=0$ should be understood
as a limiting case of (\ref{eq:Whittaker-Pi-gen}), and it is trivial.
\end{remark*}

\begin{remark}\label{rem:Whittaker-Pi--converg} Regarding the convergence
of the series, it is guaranteed by the assumption $0\leq r_{0}<r$.
Let us shortly analyze this point. Put
\[
t_{\ell}:=\left(\frac{r+r_{0}}{r\,r_{0}}\right)^{\!\mu+1/2}\,\frac{(\mu-\kappa+1/2)_{\ell}\,M_{\kappa,\ell+\mu}(r_{0})W_{\kappa,\ell+\mu}(r)}{(\ell+2\mu)_{\ell}\,\ell!\,W_{\kappa,\mu}(r+r_{0})}\,,
\]
so that (\ref{eq:Whittaker-Pi-gen}) becomes $\sum_{\ell=0}^{\infty}(-1)^{\ell}\,t_{\ell}=1$.
Referring to (\ref{eq:M-asympt}) and (\ref{eq:W-asympt}), we replace
the Whittaker functions in the numerator by their leading asymptotic
terms and get
\[
t_{\ell}=\frac{2^{2\mu-1}(r+r_{0})^{\mu+1/2}}{\sqrt{\pi}\,r^{2\mu}\,W_{\kappa,\mu}(r+r_{0})}\,\frac{(\mu-\kappa+1/2)_{\ell}\Gamma(\ell+\kappa+\mu)}{\ell!\,(\ell+2\mu)_{\ell}}\!\left(\frac{4r_{0}}{r}\right)^{\!\ell}\big(1+O(\ell^{-1})\big).
\]
Next we do the same for the factorial and the Pochhammer symbols in
the expression while using Stirling's asymptotic formula. We find
that for large $\ell$,
\[
t_{\ell}=\frac{(r+r_{0})^{\mu+1/2}}{\Gamma(\mu-\kappa+1/2)\,r^{2\mu}\,W_{\kappa,\mu}(r+r_{0})}\,\ell^{2\mu-1}\!\left(\frac{r_{0}}{r}\right)^{\!\ell}\big(1+O(\ell^{-1})\big).
\]
This obviously guarantees the convergence.

But on the other hand, the series turns out to be numerically quite
unstable for large values of $\mu$. In such a case we can be dealing
with an alternating series with many summands in its beginning attaining
huge values. Then significant cancellations of the terms necessarily
happen. From the numerical point of view this is a troublesome situation.
As an example let us consider the case with $\mu=20$, $\kappa=1$,
$r_{0}=1$, and $r=2$. The Computer Algebra System \emph{Mathematica},
as of version 14.0.0, gives the values $t_{0}=1.07239\times10^{7}$,
$t_{145}=3214.65$, and is not capable to compute $t_{\ell}$ for
higher indices. Nonetheless replacing the involved Wittaker functions
by their leading asymptotic terms for $\mu$ large one finds that
the first index for which $t_{\ell}$ attains a value smaller than
$0.1$ is $\ell=168$. This shows that this concrete series starts
to rapidly converge to its final sum only for very large summation
indices. \end{remark}

\begin{proof} In view of (\ref{eq:M-def}) and (\ref{eq:W-def})
the equation can be rewritten as
\begin{eqnarray}
 &  & \sum_{\ell=0}^{\infty}(-1)^{\ell}\,\frac{(\mu-\kappa+1/2)_{\ell}}{(\ell+2\mu)_{\ell}\,\ell!}\,(rr_{0})^{\ell}\nonumber \\
 &  & \qquad\qquad\times\,_{1}F_{1}\!\left(\ell+\mu-\kappa+\frac{1}{2};2\ell+2\mu+1;r_{0}\right)U\!\left(\ell+\mu-\kappa+\frac{1}{2},2\ell+2\mu+1,r\right)\nonumber \\
 &  & =U\!\left(\mu-\kappa+\frac{1}{2},2\mu+1,r+r_{0}\right)\!.\label{eq:MW-add-0}
\end{eqnarray}
We have \cite[Eq. 13.4.21]{AbramowitzStegun}
\[
\frac{\partial^{j}}{\partial r^{j}}U(a,b,r)=(-1)^{j}(a)_{j}U(a+j,b+j,r),\ \ j\in\ \mathbb{Z}_{+}.
\]
whence
\[
U\left(a,b,r+r_{0}\right)=\sum_{n=0}^{\infty}\frac{(-1)^{n}\,(a)_{n}}{n!}\,U(a+n,b+n,r)\,r_{0}^{\,n}.
\]
Using, in (\ref{eq:MW-add-0}), this power expansion as well as the
power expansion of $_{1}F_{1}$ we get
\begin{eqnarray*}
 &  & \sum_{\ell=0}^{\infty}(-1)^{\ell}\frac{(1/2+\mu-\kappa)_{\ell}}{(l+2\mu)_{\ell}\,\ell!}\sum_{j=0}^{\infty}\frac{(\ell+\mu-\kappa+1/2)_{j}}{(2l+2\mu+1)_{j}\,j!}\\
 &  & \qquad\times\,U\!\left(\ell+\mu-\kappa+\frac{1}{2},2\ell+2\mu+1,r\right)r^{\ell}r_{0}^{\,\ell+j}\\
 &  & =\sum_{n=0}^{\infty}\frac{(-1)^{n}\,(\mu-\kappa+1/2)_{n}}{n!}\,U\!\left(\mu-\kappa+n+\frac{1}{2},2\mu+n+1,r\right)r_{0}^{\,n}.
\end{eqnarray*}
Comparing the coefficients at the same powers of $r_{0}$ we obtain
an equivalent countable system of equations, with $n\in\mathbb{Z}_{+}$,
\begin{eqnarray*}
 &  & \sum_{\ell=0}^{n}(-1)^{\ell}\binom{n}{\ell}\frac{(2\ell+2\mu)}{(\ell+2\mu)_{n+1}}\,r^{\ell}U\!\left(\mu+\ell-\kappa+\frac{1}{2},2\mu+2\ell+1,r\right)\\
 &  & =(-1)^{n}\,U\!\left(\mu-\kappa+n+\frac{1}{2},2\mu+n+1,r\right)\!.
\end{eqnarray*}
Expressing reversely the confluent hypergeometric function $U$ in
terms of the Whittaker function $W$ we get 
\[
\sum_{\ell=0}^{n}(-1)^{\ell}\binom{n}{\ell}\frac{2\mu+2\ell}{(2\mu+\ell)_{n+1}}\,W_{\kappa,\mu+\ell}(r)=(-1)^{n}\,r^{-n/2}\,W_{\kappa-n/2,\mu+n/2}(r).
\]
This identity has been proven in Proposition~\ref{thm:W-summ-W}.
\end{proof}

Finally we prove a summation formula for the Whittaker functions $M_{\kappa,\mu}$.

\begin{proposition} For $z\in\mathbb{C}$, $\kappa\in\mathbb{C}$,
$\mu>0$ and $c\in[-1,1\,]$,
\begin{eqnarray}
 &  & \sum_{\ell=0}^{\infty}\frac{(\mu-\kappa+1/2)_{\ell}}{(2\mu)_{2\ell}}\,z^{\ell}\,_{1}F_{1}\!\left(\ell+\mu-\kappa+\frac{1}{2};2\ell+2\mu+1;z\right)C_{\ell}^{(\mu)}(c)\nonumber \\
 &  & =\,_{1}F_{1}\!\left(\mu-\kappa+\frac{1}{2};\mu+\frac{1}{2};\frac{1+c}{2}\,z\right),\label{eq:F11-summ-C}
\end{eqnarray}
or, if rewritten in terms of the Whittaker functions, with $\gamma\in[-\pi,\pi\,]$,
\begin{eqnarray}
 &  & z^{-\mu-1/2}\sum_{\ell=0}^{\infty}\frac{(\mu-\kappa+1/2)_{\ell}}{(2\mu)_{2\ell}}\,M_{\kappa,\ell+\mu}(z)\,C_{\ell}^{(\mu)}(\cos(\gamma))\nonumber \\
 &  & =e^{-z/2}\,_{1}F_{1}\!\left(\mu-\kappa+\frac{1}{2};\mu+\frac{1}{2};\cos^{2}\!\left(\frac{\gamma}{2}\right)z\right)\!.\label{eq:M-summ-gen}
\end{eqnarray}
\end{proposition}

\begin{remark} Here we tacitly assume that $z^{-\mu-1/2}M_{\kappa,\ell+\mu}(z)$
is understood as an entire function -- first defined on the positive
half-line and then admitting an unambiguous continuation to the entire
complex plane as an analytic function.

The particular case $\gamma=0$ gives
\[
\sum_{\ell=0}^{\infty}\frac{(\mu-\kappa+1/2)_{\ell}}{(\ell+2\mu)_{\ell}\,\ell!}\,M_{\kappa,\ell+\mu}(z)=e^{-z/2}\,_{1}F_{1}\!\left(\mu-\kappa+\frac{1}{2};\mu+\frac{1}{2};z\right),
\]
and the particular case $\gamma=\pi$ gives
\begin{equation}
z^{-\mu-1/2}\sum_{\ell=0}^{\infty}(-1)^{\ell}\,\frac{(\mu-\kappa+1/2)_{\ell}}{(l+2\mu)_{\ell}\,\ell!}\,M_{\kappa,\ell+\mu}(z)=e^{-z/2}.\label{M-summ-part-gamma-pi}
\end{equation}
Note that
\[
C_{\ell}^{(\mu)}(1)=\frac{(2\mu)_{\ell}}{\ell!}\,,\ C_{\ell}^{(\mu)}(-1)=(-1)^{\ell}\,\frac{(2\mu)_{\ell}}{\ell!}\,.
\]
We remark that (\ref{M-summ-part-gamma-pi}) is a generalization of
(\ref{eq:E-summ-M}). \end{remark}

\begin{remark} Convergence of the series in (\ref{eq:M-summ-gen})
can be justified by exploring the asymptotic behavior of the summands.
Regarding the Gegenbauer polynomials, an elaborate asymptotic expansion
for large orders is derived in a recent paper \cite{Dunster}. For
our purposes a comparatively simple approach is sufficient. From equation
(3.30) in \cite{Askey} one infers that
\begin{eqnarray*}
 &  & \hskip-1.5em\frac{C_{\ell}^{(\nu)}(\cos(\theta))}{C_{\ell}^{(\nu)}(1)}=\frac{2\Gamma\!\left(\nu+\frac{1}{2}\right)}{\sqrt{\pi}\,\Gamma(\nu)}\\
 &  & \hskip-1.5em\times\int_{0}^{\pi/2}\cos^{2\nu-1}(\varphi)\cos\!\left(\!\ell\arccos\!\left(\!\frac{\cos(\theta)}{\sqrt{1-\sin^{2}(\theta)\cos^{2}(\varphi)}}\right)\!\!\right)\!(1-\sin^{2}(\theta)\cos^{2}(\varphi))^{\ell/2}\,d\varphi
\end{eqnarray*}
for $\nu>0$, $0\leq\theta\leq\pi$ and $\ell\in\mathbb{Z}_{+}$.
It follows that
\[
\frac{|C_{\ell}^{(\nu)}(\cos(\theta))|}{C_{\ell}^{(\nu)}(1)}\leq\frac{2\,\Gamma\!\left(\nu+\frac{1}{2}\right)}{\sqrt{\pi}\,\Gamma(\nu)}\int_{0}^{\pi/2}\cos^{2\nu-1}(\varphi)\,d\varphi=1.
\]
Thus for $c\in[-1,1\,]$ we have
\[
|C_{\ell}^{(\mu)}(c)|\leq C_{\ell}^{(\mu)}(1)=\frac{(2\mu)_{\ell}}{\ell!}=\frac{\ell^{2\mu-1}}{\Gamma(2\mu)}\,\big(1+O(\ell^{-1})\big)\ \ \text{as}\ \ell\to\infty.
\]

Furthermore, for $\ell$ sufficiently large, surely for any $\ell$
such that $\ell+2\mu+1>|\mu-\kappa+1/2|$, we can estimate
\[
\forall n\geq0,\ \frac{|(\ell+\mu-\kappa+1/2)_{n}|}{(2\ell+2\mu+1)_{n}}\leq1,
\]
and therefore
\[
\left|\,_{1}F_{1}\!\left(\ell+\mu-\kappa+\frac{1}{2};2\ell+2\mu+1;z\right)\right|\leq K(\kappa,\mu)\,e^{|z|}
\]
where $K(\kappa,\mu)$ is a constant independent of $\ell$ and $z$
although it may depend on $\kappa$ and $\mu$. The convergence now
becomes obvious.

These estimates even show that the LHS of (\ref{eq:F11-summ-C}) is
an entire function. \end{remark}

\begin{proof} Using the power series expansion for the hypergeometric
series we have to show that
\begin{eqnarray*}
 &  & \sum_{\ell=0}^{\infty}\sum_{j=0}^{\infty}\frac{(\mu-\kappa+1/2)_{\ell}}{(2\mu)_{\ell}(l+2\mu)_{\ell}}\frac{(\ell+\mu-\kappa+1/2)_{j}}{(2\ell+2\mu+1)_{j}\,j!}\,C_{\ell}^{(\mu)}(c)\,z^{\ell+j}\\
 &  & =\sum_{n=0}^{\infty}\frac{(\mu-\kappa+1/2)_{n}}{(\mu+1/2)_{n}\,n!}\left(\frac{1+c}{2}\right)^{\!n}z^{n}.
\end{eqnarray*}
Comparing coefficients at the same powers of $z$ we get an equivalent
countable set of equations, with $n\in\mathbb{Z}_{+}$,
\[
\sum_{\ell=0}^{n}\frac{(\mu-\kappa+1/2)_{\ell}}{(2\mu)_{2\ell}}\,\frac{(\ell+\mu-\kappa+1/2)_{n-\ell}}{(2\ell+2\mu+1)_{n-\ell}(n-\ell)!}\,C_{\ell}^{(\mu)}(c)=\frac{(\mu-\kappa+1/2)_{n}}{(\mu+1/2)_{n}\,n!}\left(\frac{1+c}{2}\right)^{n}.
\]
These equations can be simplified using straightforward manipulations,
\[
\sum_{\ell=0}^{n}\frac{(2\mu+2\ell)}{(2\mu)_{n+\ell+1}(n-\ell)!}\,C_{\ell}^{(\mu)}(c)=\frac{2^{n}(\mu)_{n}}{(2\mu)_{2n}\,n!}\,(1+c)^{n}.
\]
Recall that \cite[Eq. (9.8.19)]{KoekoekLeskySwarttouw}
\[
C_{\ell}^{(\mu)}(c)=\frac{(2\mu)_{\ell}}{\ell!}\,\,_{2}F_{1}\!\left(-\ell,2\mu+\ell;\mu+\frac{1}{2};\frac{1-c}{2}\right).
\]
After the substitution $c=1-2w$ we get the equation
\[
\sum_{\ell=0}^{n}\frac{(2\mu+2\ell)}{(2\mu)_{n+\ell+1}(n-\ell)!}\,\frac{(2\mu)_{\ell}}{\ell!}\,\,_{2}F_{1}\!\left(-\ell,2\mu+\ell;\mu+\frac{1}{2};w\right)=\frac{2^{2n}(\mu)_{n}}{(2\mu)_{2n}n!}\,(1-w)^{n}.
\]
And after the power series expansion of the hypergeometric function
and interchanging the order of summation we obtain
\[
\sum_{j=0}^{n}\left(\sum_{\ell=j}^{n}\frac{(2\mu+2\ell)}{(2\mu)_{n+\ell+1}(n-\ell)!}\,\frac{(2\mu)_{\ell}}{\ell!}\,\frac{(-\ell)_{j}(2\mu+\ell)_{j}}{(\mu+1/2)_{j}j!}\right)\!w^{j}=\frac{2^{2n}(\mu)_{n}}{(2\mu)_{2n}n!}\,(1-w)^{n}.
\]
Comparing coefficients at the same powers of $w$ leads to the equations
\[
\sum_{\ell=j}^{n}\frac{(2\mu+2\ell)}{(2\mu)_{n+\ell+1}(n-\ell)!}\frac{(2\mu)_{\ell}}{\ell!}\frac{(-\ell)_{j}(2\mu+\ell)_{j}}{(\mu+1/2)_{j}}=\frac{(-1)^{j}}{(n-j)!}\,\frac{2^{2n}(\mu)_{n}}{(2\mu)_{2n}},\ \ 0\leq j\leq n,
\]
and this can be further rewritten,
\[
\sum_{\ell=j}^{n}\binom{n-j}{\ell-j}\frac{2\mu+2\ell}{(2\mu+\ell+j)_{n-j+1}}=\frac{(\mu+1/2)_{j}}{(\mu+1/2)_{n}},\ \ 0\leq j\leq n.
\]
Shifting the summation index $\ell\to\ell+j$ and applying the substitutions
$n=N+j$, $\nu=\mu+j$, we obtain an identity (equation (\ref{eq:L}))
which is proven in Lemma~\ref{thm:L} below thus concluding this
proof. \end{proof}

\begin{lemma}\label{thm:L} For $N\in\mathbb{Z}_{+}$ and $\nu\in\mathbb{C}$,
$\nu\neq0,-1,-2,\ldots$,
\begin{equation}
\sum_{\ell=0}^{N}\binom{N}{\ell}\frac{2\nu+2\ell}{(2\nu+\ell)_{N+1}}=\frac{1}{(\nu+1/2)_{N}}\,.\label{eq:L}
\end{equation}
\end{lemma}

\begin{proof} For $N\in\mathbb{Z}_{+}$, consider the expression
\begin{equation}
\sum_{\ell=0}^{N}\binom{N}{\ell}\frac{(2\nu+2\ell)(\nu+1/2)_{N}}{(2\nu+\ell)_{N+1}}\label{eq:L-expr}
\end{equation}
which is regarded as a function of $\nu\in\mathbb{C}$. This is a
meromorphic function on $\mathbb{C}$, and its limit, as $\nu\to\infty$,
equals 
\begin{equation}
\sum_{\ell=0}^{N}\binom{N}{\ell}\frac{1}{2^{N}}=1.\label{eq:L-expr-limit}
\end{equation}
The poles are located at the points $\nu=0,-1/2,-1,\ldots,-(2N-1)/2,-N$,
and all of them are of first order. 

Owing to the factor $(\nu+1/2)_{N}$ the singularities are removable
for
\[
\nu=-1/2,-3/2,\ldots,-(2N-1)/2.
\]
Hence it suffices to check the poles at $\nu=0,-1,\ldots,-N$. The
residue at a pole $\nu=-t$, $0\leq t\leq N$, equals
\begin{eqnarray*}
 &  & (-t+1/2)_{N}\sum_{\ell=\max\{2t-N,0\}}^{\min\{2t,N\}}\binom{N}{\ell}\frac{-2t+2\ell}{(-2t+\ell)_{2t-\ell}(1)_{N-2t+\ell}}\\
 &  & =(-t+1/2)_{N}\sum_{\ell=\max\{2t-N,0\}}^{\min\{2t,N\}}(-1)^{\ell}\binom{N}{\ell}\frac{-2t+2\ell}{(2t-\ell)!(N-2t+\ell)!}.
\end{eqnarray*}
Further we omit the nonzero factor $2\,(-t+1/2)_{N}$. Shifting the
summation index $\ell\to t+\ell$, for $t\geq N/2$ we get
\[
\sum_{l=2t-N}^{N}(-1)^{\ell}\binom{N}{\ell}\frac{-t+\ell}{(2t-\ell)!(N-2t+\ell)!}=\frac{(-1)^{t}}{N!}\sum_{\ell=t-N}^{N-t}(-1)^{\ell}\binom{N}{t+\ell}\binom{N}{t-\ell}\,\ell=0.
\]
Just note that the summands are odd in $\ell$. For $t\leq N/2$ we
similarly get
\[
\sum_{\ell=0}^{2t}(-1)^{\ell}\binom{N}{\ell}\frac{-t+\ell}{(2t-\ell)!(N-2t+\ell)!}=\frac{(-1)^{t}}{N!}\sum_{\ell=-t}^{t}(-1)^{\ell}\binom{N}{t+\ell}\binom{N}{t-\ell}\ell=0.
\]
Hence all singularities are removable and therefore expression (\ref{eq:L-expr})
equals $1$ identically from (\ref{eq:L-expr-limit}) and Liouville's
Theorem. \end{proof}

\section*{{\large{}ACKNOWLEDGMENTS}}

The author acknowledges partial support by European Regional Development
Fund Project \textquotedblleft Center for Advanced Applied Science\textquotedblright{}
No. CZ.02.1.01/0.0/0.0/16\_019/0000778. The author is indebted to
the reviewer for numerous comments helpful in improving the quality
of the paper, and in particular for Remark \ref{rem:Whittaker-Pi--converg}.


\begin{thebibliography}{10}
\bibitem{AbramowitzStegun} M.~Abramowitz, I.~A.~Stegun: \emph{Handbook
of Mathematical Functions with Formulas, Graphs, and Mathematical
Tables}, (Dover Publications, New York, 1972).

\bibitem{Askey} R.~Askey: \emph{Orthogonal Polynomials and Special
Functions}, (SIAM, Philadelphia, 1975).

\bibitem{Blinder-density} S.~M.~Blinder: \emph{Generalized Unsöld
theorem and radial distribution function for hydrogenic orbitals},
J.~Math. Chem., \textbf{14} (1993), 319-324.

\bibitem{DLMF} \emph{NIST Digital Library of Mathematical Functions}.
https://dlmf.nist.gov/, Release 1.1.11 of 2023-09-15. F. W. J. Olver,
A. B. Olde Daalhuis, D. W. Lozier, B. I. Schneider, R. F. Boisvert,
C. W. Clark, B. R. Miller, B. V. Saunders, H. S. Cohl, and M. A. McClain,
eds.

\bibitem{Dunster-MW} T.~M.~Dunster: \emph{Uniform asymptotic expansions
for the Whittaker functions $M_{\kappa,\mu}(z)$ and $W_{\kappa,\mu}(z)$
with $\mu$ large}, Proc.~R.~Soc.~A \textbf{477} (2021), 0360.

\bibitem{Dunster} T.~M.~Dunster: \emph{Uniform asymptotic expansions
for Gegenbauer polynomials and related functions via differential
equations having a simple pole}, Constr. Approx. (2023), https://doi.org/10.1007/s00365-023-09645-1.

\bibitem{ErdelyiMagnusOberhettingerTricomi} A.~Erdélyi, W.~Magnus,
F.~Oberhettinger, F.~G.~Tricomi: \emph{Higher Transcendental Functions.
Vol. II}. (McGraw-Hill Book Company, Inc., New York-Toronto-London,
1953).

\bibitem{Ferrers} N.~M.~Ferrers: \emph{An elementary treatise on
spherical harmonics and subjects connected with them}, (MacMillan,
London, 1877).

\bibitem{GradshteynRyzhik} I.~S.~Gradshteyn, I.~M.~Ryzhik: \emph{Table
of Integrals, Series, and Products}, Edited by A.~Jeffrey and D.~Zwillinger,
(Academic Press, Amsterdam, 2007).

\bibitem{Hostler} L.~C.~Hostler: \emph{Coulomb Green\textquoteright s
function in closed form}, Bull. Am. Phys. Soc. \textbf{7} (1962),
609.

\bibitem{HostlerPratt} L.~C.~Hostler, R.~H.~Pratt: \emph{Coulomb
Green's function in closed form}, Phys. Rev. Lett. \textbf{10} (1963),
469-470.

\bibitem{KoekoekLeskySwarttouw} R.~Koekoek, P.~A.~Lesky, R.~F.~Swarttouw:\textit{
Hypergeometric Orthogonal Polynomials and Their $q$-Analogues}, (Springer,
Berlin, 2010).

\bibitem{MalecekNadenik} K.~Male\v{c}ek, Z.~N\'aden\'{\i}k: \emph{On
the inductive proof of Legendre addition theorem}, Studia Geophysica
et Geodaetica \textbf{45} (2001), 1-11.

\bibitem{Olver} F.~W.~J.~Olver: \emph{Asymptotics and Special
Functions}, (A.~K.~Peters, Wellesley, 1997). 

\bibitem{Rauh} A.~Rauh: \emph{On the singularities of the nonrelativistic
Coulomb Green\textquoteright s function}, Adv. Stud. Theor. Phys.
\textbf{13} (2019), 175-187.

\bibitem{VasanSeetharaman} S.~S.~Vasan, M.~Seetharaman: \emph{The
Coulomb Green's function revisited}, Pramana -- J.~Phys. \textbf{45}
(1995), 165-174.

\bibitem{Watson} G.~N.~Watson: \emph{A Treatise on the Theory of
Bessel Functions}, 2nd edition, (Cambridge University Press, Cambridge,
1944).
\end{thebibliography}
\end{document}